\documentclass[prx,twocolumn,aps,superscriptaddress,eqsecnum,floatfix]{revtex4-1}
\usepackage[utf8]{inputenc}
\usepackage{amsmath}
\usepackage{amssymb}
\usepackage{graphicx}
\usepackage{color}
\usepackage{float}
\usepackage[bookmarks=false,linkcolor=blue,urlcolor=blue,colorlinks,citecolor=blue]{hyperref}

\usepackage{bm}
\usepackage{bbm}

\begin{document}
\title{Phase diagram and phonon-induced backscattering in  topological insulator nanowires}

\author{Kathrin Dorn,$^1$ Alessandro De Martino,$^2$ and Reinhold Egger}
\affiliation{Institut f\"ur Theoretische Physik, Heinrich-Heine-Universit\"at, 40225  D\"usseldorf, Germany\\
$^2$~Department of Mathematics, City, University of London, EC1V 0HB London, UK}

\begin{abstract}
We present an effective low-energy theory of electron-phonon coupling effects 
for clean cylindrical topological insulator nanowires.  
Acoustic phonons are modelled by isotropic elastic continuum theory with 
stress-free boundary conditions.  We take into account the deformation potential coupling between phonons and
helical surface Dirac fermions,   and also include electron-electron interactions within the bosonization approach. 
For half-integer values of the magnetic flux $\Phi_B$ along the wire,
the low-energy theory admits an exact solution since a topological protection mechanism then rules out phonon-induced $2k_F$-backscattering processes. 
We determine the zero-temperature phase diagram and identify
a regime dominated by superconducting pairing of surface states. As example, we consider the phase diagram of HgTe nanowires. 
We also determine the phonon-induced electrical resistivity, where we find a quadratic dependence on the flux deviation $\delta\Phi_B$ 
 from the nearest half-integer value.
\end{abstract}
\date{\today}
\maketitle

\section{Introduction}\label{sec1}

The spin-momentum locked Dirac fermion surface states of three-dimensional (3D) topological insulators (TIs) 
\cite{Fu2007,Qi2008,Hasan2010,Liu2010,Qi2011,Ando2013} have been intensely studied over the past decade. 
For several established TI materials, phonon-induced effects have been examined for various physical observables, e.g., 
the temperature-dependent quasi-particle lifetime measured from the linewidth in 
angle-resolved photoemission spectroscopy (ARPES), or the electrical 
resistivity obtained from transport experiments, cf.~Refs.~\cite{Zhu2011,Hatch2011,Zhu2012,Wang2012,Kim2012,Pan2012,Pan2013,Chen2013,Kondo2013,Crepaldi2013,Howard2014,Costache2014,Sobota2014,Ando2014,Glinka2015,Tamtogl2017,Jia2017,Wiesner2017}.  On the theoretical side, substantial progress has also been made, see, e.g., 
Refs.~\cite{Huang2008,Thalmeier2011,Giraud2011,Giraud2012,Budich2012,Garate2013,Zhang2013,Parente2013,Sarma2013,Wang2014,Heid2017}.
Here the effective four-band model for the low-energy electronic structure of 3D TIs \cite{Liu2010} provides a convenient starting point for
 analytical studies, where surface states follow by imposing Dirichlet boundary conditions at the surface.  
Apart from theoretical studies of phonon-induced effects for the half-space \cite{Giraud2011,Budich2012,Garate2013,Parente2013,Sarma2013} and thin-film  \cite{Giraud2012} TI geometries, 
surface states have also been constructed for one-dimensional (1D) TI nanowires 
\cite{Ran2009,Ostrovsky2010,Zhang2010,Bardarson2010,Egger2010,Kundu2011,Bardarson2013}, typically by assuming a cylindrical cross section to simplify  calculations.

Measurements of the quasi-particle lifetime of surface states in 3D TIs can give precious information on the 
electron-phonon coupling strength. For instance, the phonon dispersion relation can be measured using coherent
helium beam surface scattering, where the observation of a Kohn anomaly may allow one to extract the coupling constant. 
(However, this method may encounter difficulties for acoustic phonons \cite{Zhu2011,Zhu2012}.) Coupling values for acoustic phonons have
also been reported from other experimental approaches such as ARPES, mostly for TIs belonging to the Bi$_2$Se$_3$ family \cite{Hatch2011,Wang2012,Kim2012,Pan2012,Pan2013,Chen2013,Kondo2013,Crepaldi2013,Howard2014,Costache2014,Sobota2014,Ando2014,Glinka2015,Tamtogl2017,Jia2017,Wiesner2017}.
A rather wide discrepancy between the reported values exists which (at least partially) may be due to additional optical phonon effects, see, e.g., Ref.~\cite{Kondo2013}.  
The remaining differences can likely be explained by noting that different experiments were carried out at
different Fermi energies, sample qualities, and/or  temperatures.  All these parameters can strongly influence the expected value of the coupling constant  
\cite{Budich2012,Sarma2013}.  Most likely, the electron-phonon coupling between surface states and acoustic phonons is 
of intermediate strength for TIs in the Bi$_2$Se$_3$ family. 

In this paper, we investigate phonon-induced effects in cylindrical TI nanowires, taking into account the deformation potential coupling between surface Dirac fermion states and 
acoustic phonons, electron-electron interactions, as well as the presence of a dimensionless magnetic flux $\Phi_B$ (in units of the flux quantum) piercing the nanowire.
The deformation potential is expected to be the dominant coupling mechanism, affecting both the low-temperature resistivity and the linewidth of
photoemission peaks. Following Refs.~\cite{Giraud2011,Giraud2012}, acoustic phonons are here modelled in terms of isotropic elastic continuum theory \cite{Landau7}.
In view of the complex quintuple-layer structure of Bi$_2$Se$_3$ or Bi$_2$Te$_3$, the validity of the isotropic continuum approach
with just two elastic Lam\'e constants may come as a surprise. Nonetheless, the usefulness of
isotropic elastic continuum theory for modelling low-energy phonons for bulk 3D TI materials has
been established by ab-initio studies \cite{Huang2008}. 
Using stress-free boundary conditions at the sample boundaries, the uncoupled phonon eigenmodes can then be
determined for the geometry of interest. Thereby one also obtains the electron-phonon coupling
Hamiltonian from the deformation potential. Below we define a dimensionless electron-phonon coupling 
parameter $A$, see Eq.~\eqref{bdef}, to quantify the electron-phonon interaction strength.
We note in passing that for rectangular quantum wires described by the Schr\"odinger equation, the coupled electron-phonon
problem has been studied in Ref.~\cite{Svizhenko1998}. However, for the helical Dirac surface states of interest below, the physics 
turns out to be rather different.  For instance, in our system phonon-induced $2k_F$-backscattering is forbidden
for half-integer  $\Phi_B$.

The probably most important restriction of our theoretical approach comes from the neglect of disorder.  
Magneto-transport measurements for TI nanowires have been performed by different groups \cite{Dufouleur2013,Hong2014,Cho2015,Ziegler2018,Munning2019}.   
While most of the published experimental results need to invoke disorder effects for a consistent explanation, 
ballistic (basically disorder free) experiments for TI nanowires have also been reported \cite{Cho2015}.  In addition, the TI material HgTe has emerged as  
a particularly clean platform \cite{Capper2011,Jain2013}, with nanowire experiments being already available \cite{Ziegler2018}.  Moreover, direct evidence for Dirac surface state subbands in  (Bi$_{1-x}$Sb$_x)_2$Te$_3$-based TI nanowires  has recently been obtained from transport experiments \cite{Munning2019}. We note in passing that albeit those TI nanowires have a hexagonal cross-section, the observed Dirac subbands essentially follow analytical predictions obtained for cylindrical nanowires, see Ref.~\cite{Munning2019}.
While we neglect disorder, we will include \emph{electron-electron interactions} on a non-perturbative level within the helical Luttinger liquid framework \cite{Egger2010,Gogolinbook}.
Note that electron-electron interactions may also give temperature-dependent contributions to the surface resistivity, as described for a semi-infinite TI geometry in Ref.~\cite{Pal2012}.
In terms of the helical Luttinger liquid framework, interactions are encoded by a parameter $K$.  The value $K=1$ describes the noninteracting case, while for repulsive Coulomb interactions, one finds $K<1$.  The gapped phonon branches neglected in our analysis below could be included by a renormalization of $K$ towards larger values \cite{Gogolinbook}.  
In that way, also the regime $K>1$ may become reachable.

\begin{figure}[t]
\includegraphics[width=0.52\textwidth]{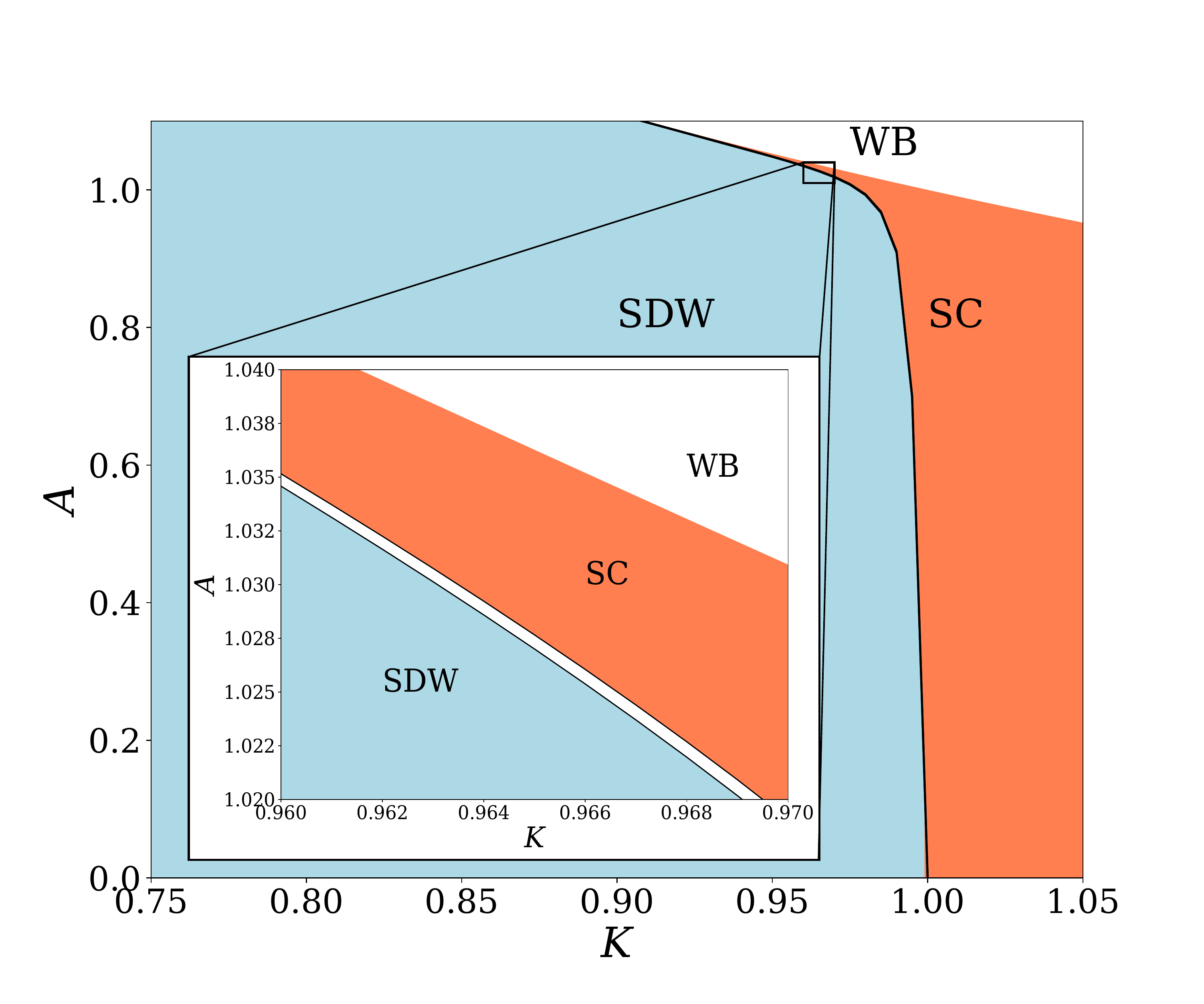}
\caption{Zero-temperature phase diagram of a TI nanowire (pierced by half-integer flux $\Phi_B$) in the $K$-$A$ plane, where $A$  in Eq.~\eqref{bdef} parametrizes the electron-phonon coupling strength and $K$ is the Luttinger liquid parameter, encoding the effective electron-electron interaction strength.  We use material parameters for HgTe, see main text.  For $A\ge 1/K$, the system is unstable (Wentzel-Bardeen regime, `WB'). For $A<1/K$, three phases are possible: Superconducting correlations dominate in the `SC' part of the phase diagram.  A spin-density wave phase (with ordering along the nanowire axis) 
is denoted by `SDW'.    The inset gives a magnified view of a region where
the tiny intermediate `metallic' phase (white) is visible. Here conventional 1D current-current correlations represent the slowest decay. For details, see Sec.~\ref{sec5}.
}
\label{fig1}
\end{figure}

After the construction of the low-energy theory for the interacting electron-phonon system in a TI nanowire, we will use this theory to discuss two different physical questions.
First, we will study the zero-temperature \emph{phase diagram} for half-integer $\Phi_B$. Following standard practice  \cite{Gogolinbook}, a `phase' will here be identified by the slowest algebraic
decay of order parameter correlations.  We will, in particular, search for superconducting instabilities of the surface Dirac fermions
due to the effectively attractive interactions mediated by acoustic phonons.  Previous theoretical works have studied related questions for the semi-infinite TI geometry:
While the authors of Ref.~\cite{Sarma2013} have argued that surface superconductivity appears below $T_c\approx 1.3$~K,  complementary work
found a parametrically smaller critical temperature $T_c \alt 1$~mK \cite{Parente2013}.  
Similar issues have also been studied for standard (non-helical) Luttinger liquids arising, e.g., in carbon nanotubes, see Refs.~\cite{Loss1992,DeMartino2003,Schulz2010}.   
The issue of intrinsic superconductivity in TI nanowires is also of relevance for the possible existence of Majorana bound states in such systems
\cite{Cook2011,Cook2012,Fidkowski2011,Manousakis2017}. Such states are highly promising candidates for topological quantum information processing schemes. 
Our nonperturbative analysis of the zero-temperature phase diagram for half-integer $\Phi_B$ arrives at the scenario shown in Fig.~\ref{fig1}, where we 
depict the dominant order parameter correlations in the $K$-$A$ plane.   In particular, for material parameters corresponding to HgTe nanowires, we thereby 
can identify the parameter regions where superconducting pairing of surface states is expected to be dominant.  We provide a detailed derivation 
and discussion of the phase diagram in Sec.~\ref{sec5}.

As second application, we will study the \emph{phonon-induced electrical resistivity} for $\Phi_B$ close to half-integer values.
For precisely half-integer $\Phi_B$, the topological protection of surface states  \cite{Ran2009,Ostrovsky2010,Zhang2010,Bardarson2010,Egger2010,Kundu2011,Bardarson2013} implies the absence of phonon-induced $2k_F$ backscattering.  As a consequence, the resistivity of a disorder-free interacting TI nanowire vanishes even in the presence of electron-phonon coupling.  However, once the flux $\Phi_B$ is tuned away from half-integer values, we will see that a finite resistivity emerges at $T\ne 0$. 
The specific predictions for the $T$ dependence and for the dependence on $\Phi_B$ made below should allow for direct experimental tests of our theory.
For conventional 1D quantum wires, phonon-induced conductance corrections have been studied theoretically 
in Refs.~\cite{Voit1987,Bockelmann1990,Shik1993,Mickevicius1993,Gurevich1995,Gurevich1995,Gurevich1995b,Seelig2005,Yurkevich2013}.

The remainder of this paper is organized as follows.   We describe the surface states of a TI nanowire in Sec.~\ref{sec2}, 
where we also include electron-electron interactions within the Abelian bosonization approach \cite{Gogolinbook}. 
For modelling the acoustic phonon spectrum, in Sec.~\ref{sec3} we start from isotropic elastic continuum theory for an infinitely long
cylindrical wire. Although this problem has been studied long ago \cite{Pochhammer1876,Chree1889,Love,Graff},  
in order to keep the paper self-contained, we present an independent derivation.
By  employing angular momentum eigenstates, our formulation also yields particularly transparent expressions.  
In Sec.~\ref{sec4}, as the dominant electron-phonon coupling term, we take into account the deformation potential and analyze the resulting 
low-energy theory for the coupled electron-phonon system.
In Sec.~\ref{sec5}, we derive the phase diagram in Fig.~\ref{fig1} which applies to TI nanowires pierced by  half-integer flux $\Phi_B$.
Next, in Sec.~\ref{sec6}, we discuss the phonon-induced electric resistivity for $\Phi_B$ close to half-integer values but in the absence of 
electron-electron interactions (corresponding to $K=1$).  We use the Boltzmann equation approach by Gurevich \textit{et al.}~\cite{Gurevich1995} to 
obtain the temperature-dependent resistivity in the linear response regime.
Finally, we offer some concluding remarks in Sec.~\ref{sec7}. Throughout the paper, we use units with $\hbar=k_B=1$.

\section{Electronic surface states of topological insulator nanowires}\label{sec2}

We first describe our model for the low-energy electronic states of a cylindrical TI nanowire.  
Typically, TI materials are characterized by a sizeable bulk gap of order $\Delta_b\simeq 0.3$~eV \cite{Fu2007,Qi2008,Hasan2010,Liu2010,Qi2011}. As long as the Fermi energy resides well within the bulk gap and provided that one has sufficiently clean materials to realize the ballistic transport regime, only surface states will be relevant for the low-energy transport properties.
For a cylindrical wire of radius $R$, the electronic surface spectrum consists of massive 1D Dirac fermion modes with conserved momentum $k$ along the cylinder axis ($\hat e_z$)  \cite{Ran2009,Ostrovsky2010,Zhang2010,Bardarson2010,Egger2010,Kundu2011,Bardarson2013}.  Below we include an axial magnetic field $B$ giving rise to the dimensionless flux $\Phi_B=\pi R^2 B/(h/e)$ piercing the nanowire. 

In cylindrical coordinates ($r,\phi,z)$, with  unit vectors $(\hat e_r,\hat e_\phi,\hat e_z)$,
the electronic single-particle Hamiltonian describing surface states with conserved momentum $k$ 
is a Dirac Hamiltonian wrapped onto the cylinder surface \cite{Ran2009,Egger2010,Kundu2011},
\begin{equation}\label{heff}
    {\cal H}_{\rm el}(k) =  e^{-i\phi\sigma_z/2} \left( v_1 k \sigma_y - \frac{v_2}{R}(-i\partial_\phi+\Phi_B)\sigma_z
    \right) e^{i\phi\sigma_z/2},
\end{equation} 
with the Fermi velocities $v_1$ ($v_2$) along $\hat e_z$ (perpendicular to $\hat e_z$). The
 Pauli matrices $\sigma_{x,y,z}$ act in spin space.  The dispersion relation of the 1D fermion modes is thus given by ($\pm$ refers to conduction and valence bands) \cite{Ran2009,Ostrovsky2010,Bardarson2010}
\begin{equation}\label{ener2}
    E_{j,\pm}(k) = \pm \sqrt{v^2_1k^2+v_2^2(j+\Phi_B)^2/R^2}.
\end{equation}
Different bands are distinguished by the half-integer eigenvalue $j$ of the conserved $z$-component of the total angular momentum operator. 

We next note that for integer $\Phi_B$, a time-reversal (${\cal T}$) transformation connects the states $(k,j+\Phi_B)\leftrightarrow (-k,-j-\Phi_B)$. Due to this emergent ${\cal T}$-symmetry, all states are arranged into doubly degenerate Kramers pairs.  
While elastic scattering between such pairs is forbidden by virtue of the Kramers theorem, $2k_F$ backscattering ($k\to -k$) for given $j$ is allowed and there is no protection against elastic disorder effects.
However, for \emph{half-integer} $\Phi_B$,  the emergent
${\cal T}$-symmetry now comes with a topological protection against weak spin-conserving backscattering. This is because for the special massless 1D Dirac mode with $j=-\Phi_B$, the two states with momentum $\pm k$ constitute a protected Kramers pair \cite{Ran2009,Ostrovsky2010,Zhang2010,Bardarson2010}. While this scenario --- a single Dirac fermion species protected by an emergent ${\cal T}$-symmetry ---
is ruled out for conventional systems by the  Nielsen-Ninomiya theorem \cite{Hasan2010}, it can be realized using the surface states of TI nanowires with half-integer flux $\Phi_B$. 
The special mode with $j=-\Phi_B$ is protected against elastic disorder effects and dominates the physics on energy scales below $E_g\simeq v_2/R$.
On higher energy scales, also other transverse bands with $j\ne -\Phi_B$  have to be included in the theory.

We consider the case of low energies, $|E|\ll \Delta_g\equiv {\rm min}(E_g,\Delta_b)$, throughout this paper. 
Putting $\Phi_B$ to an half-integer value, we then keep only the gapless Dirac mode with $j=-\Phi_B$ and $E_{\pm}(k)=\pm v_1|k|$.  
The electron field operator is now represented in terms of the spinor  \cite{Egger2010}
\begin{equation}\label{1dexp}
    \Psi_{\rm el}(r,\phi,z) = \frac{f_\perp(r)}{\sqrt{4\pi}}  \sum_{\nu=\pm}
    e^{i\nu k_F z} e^{i(j-1/2)\phi} \psi_\nu(z) \left( \begin{array}{c} \nu \\ ie^{i\phi} \end{array}\right)
\end{equation}
with Fermi momentum $k_F=\mu/v_1$. For simplicity, we shall assume that the chemical potential is within the range $0<\mu\ll\Delta_g$. 
The slowly varying 1D fermion operators, $\psi_{\nu=+/-}(z)$, correspond to right- and left-movers, respectively.  The radial part,
$f_{\perp}(r)$, with normalization $\int_0^\infty r dr |f_{\perp}(r)|^2=1$, vanishes for $r>R$ and decays exponentially away from the surface for $r<R$. Here the radial width, $\xi_\perp$, of 
the surface state depends on microscopic details \cite{Egger2010}. For $\xi_\perp\ll R$, we have
\begin{equation}\label{fperp}
    f_{\perp}(r) \simeq 
    \sqrt{\frac{2}{\xi_\perp R}}\,
 e^{-(R-r)/\xi_\perp} \,\Theta(R-r),
\end{equation}
with the Heaviside step function $\Theta$.  

Using the standard bosonization approach, we next express the 1D field operators appearing in Eq.~\eqref{1dexp} in terms of the dual boson field operators $\theta(z)$ and $\varphi(z)$ \cite{Gogolinbook},
\begin{equation}
\label{1Dop}    \psi_{\nu=\pm}(z) \simeq \frac{1}{\sqrt{2\pi\xi_\perp}}
    e^{i\sqrt{\pi}[\varphi(z)+\nu\theta(z)]},
\end{equation}
where we identify the short-distance cutoff length with $\xi_\perp$.  Using Eq.~\eqref{1dexp}, the electron density operator is then given by
\begin{equation} \label{densop}
    \rho_{\rm el}({\bf r}) \equiv \Psi^\dagger_{\rm el}\Psi^{}_{\rm el}= \frac{1}{\sqrt{4\pi^3}} |f_\perp(r)|^2
   \,  \partial_z \theta(z).
\end{equation}
We emphasize that the standard $2k_F$-term in the density operator is not present for this topological band \cite{Egger2010}.  This fact implies that charge density wave ordering is not possible.  However, once $\Phi_B$ deviates from half-integer values, we will see in Sec.~\ref{sec6b} that a $2k_F$-oscillatory term appears in the density operator since backscattering is now allowed.

The bosonization approach is particularly advantageous for 1D systems because it allows one to easily take into account Coulomb interaction effects \cite{Gogolinbook}.
Including the dominant long-range interactions within the helical Luttinger liquid picture of Ref.~\cite{Egger2010}, the effective low-energy Hamiltonian for the many-electron system with half-integer flux $\Phi_B$ is given by a noninteracting boson theory,
\begin{equation}\label{helec}
H_{\rm el} = \frac{v_1}{2} \int dz \left[ (\partial_z \varphi)^2+ K^{-2} (\partial_z \theta)^2\right],
\end{equation}
where the Luttinger liquid parameter $K$ takes into account the effect of electron-electron interactions.  (For detailed expressions of $K$ in terms of
microscopic details, see Ref.~\cite{Egger2010}.)
The noninteracting limit corresponds to $K=1$, and repulsive interactions imply $K<1$. For instance, $K\approx 0.5$ has been estimated for Bi$_2$Se$_3$ or Bi$_2$Te$_3$ nanowires assuming that there is no closeby metallic gate \cite{Egger2010}.

\section{Acoustic phonon modes }\label{sec3}

In this section, we construct the acoustic phonon modes of a cylindrical wire by assuming that its elastic properties can be described as isotropic continuum \cite{Giraud2011,Giraud2012}.

\subsection{3D isotropic elastic continuum}

We start by considering a 3D isotropic elastic continuum described by the linearized strain tensor, $u_{ij}({\bf r})=(\partial_i u_j+\partial_j u_i)/2,$ with $i,j=x,y,z$ 
and the displacement field ${\bf u}({\bf r},t)$.
The elastic free energy density then reads \cite{Landau7}
\begin{equation}\label{freedens}
{\cal F}[u]=  \frac{\lambda}{2} \left(\text{Tr} u\right)^2 + \mu \text{Tr}(u^2),
\end{equation}
with the Lam\'e constants $\lambda$ and $\mu$, and
the stress tensor takes the form 
\begin{equation}\label{stress}
\sigma_{ij} = \lambda  \text{Tr}( u)\, \delta_{ij}  + 2\mu \, u_{ij}.
\end{equation}
From Eq.~\eqref{freedens}, the equations of motion are given by
\begin{equation} \label{wave1}
\ddot {\bf u}=
c_t^2 \Delta {\bf u} + (c_l^2-c_t^2) \nabla (\nabla \cdot {\bf u}),
\end{equation}
with the velocities for transverse,
$c_t=\sqrt{\mu/\rho_M}$, and longitudinal,
$c_l=\sqrt{(\lambda+2\mu)/\rho_M}$, sound waves. Here
$\rho_M$ is the mass density.  For  Bi$_2$Te$_3$, the isotropic elastic continuum approximation is expected to work reasonably well and one finds $\rho_M\simeq 7860$~kg$/$m$^3$, $c_t\simeq 1600$~m/s, and $c_l\simeq 2800$~m/s
\cite{Giraud2011}, cf.~Refs.~\cite{Huang2008,Jenkins1972}.  For later use, we also define the dimensionless ratio 
\begin{equation}\label{xidef}
\xi=c_t/c_l<1.
\end{equation}

The displacement field can always be represented as sum of  longitudinal and transverse parts,
\begin{equation}
{\bf u}({\bf r},t) = {\bf u}_l + {\bf u}_t=
\nabla \Phi+ \nabla \times {\bm \Psi},
\end{equation}
with a scalar potential $\Phi({\bf r},t)$ and a vector potential ${\bm \Psi}({\bf r},t)$, where  Eq.~\eqref{wave1} implies decoupled
wave equations,
\begin{equation}\label{wave2}
\left(\partial_t^2 - c^2_l \Delta\right) \Phi=0, \quad
\left(\partial_t^2 -  c^2_t \Delta\right) {\bm \Psi}=0. 
\end{equation} 
However,  boundary conditions will generally couple both potentials.
We next write Eq.~\eqref{wave2} in cylindrical coordinates.   
Translation invariance along $\hat e_z$ (we assume periodic boundary 
conditions with length $L$ and eventually let $L\to \infty$)
implies the ($z,t)$-dependence  
$\Phi,{\bm \Psi} \sim e^{i (q z- \Omega t)}$,  where
 $q$ is a conserved wave number along $\hat e_z$ and $\Omega>0$ a possible eigenfrequency.
 For convenience, we define the two wave numbers
\begin{equation}\label{klkt}
k_l=\sqrt{\frac{\Omega^2}{c_l^2}-q^2},\quad k_t=\sqrt{\frac{\Omega^2}{c_t^2}-q^2}.
\end{equation}
Second, to exploit rotation symmetry around $\hat e_z$, we expand $\Phi$ and ${\bm\Psi}$ in terms of eigenstates of  the conserved total 
angular momentum operator $J_z^{\rm ph}$. This operator  
has integer eigenvalues denoted by $m$. 

For the \emph{longitudinal part}, ${\bf u}_l=\nabla\Phi$, we  observe that $J_z^{\rm ph}$ acts like $-i\partial_\phi$ on $\Phi$. Solutions to Eq.~\eqref{wave2} are of the form
\begin{equation}
\Phi({\bf r},t)= f(r) \frac{e^{im\phi}}{\sqrt{2\pi}} \frac{e^{i q z}}{\sqrt{L}} e^{-i \Omega t},
\end{equation}
with a radial Bessel equation for $f(r)$, 
\begin{equation}\label{radeq}
\left( \frac{1}{r} \partial_r (r\partial_r)  -\frac{m^2}{r^2}+ k_l^2 \right)f(r)=0.
\end{equation}
The general solution of Eq.~\eqref{radeq} is given by
\begin{equation}
f(r)=A_1J_m(k_lr) +A_2 Y_m(k_l r),
\end{equation}
with $k_l$ in Eq.~\eqref{klkt}, arbitrary constants $A_{1,2}$, and the Bessel functions $J_m(z)$ and $Y_m(z)$ \cite{DLMF}.
Regularity at $r\to 0$ imposes $A_2=0$ unless one considers a hollow cylinder. 
After straightforward algebra we obtain, for given $(m,q)$, the longitudinal part of the displacement field as
\begin{eqnarray}\label{ulfinal}
{\bf u}_l({\bf r},t) & =& a\Bigl[
 k_l J'_{m}(k_lr)  \hat e_r  +   \frac{im}{r} J_{m}(k_lr) \hat e_\phi\\
&+&  iq J_m(k_lr)\hat e_z \Bigr] \frac{e^{im\phi}}{\sqrt{2\pi}} \frac{e^{i q z}}{\sqrt{L}}  e^{-i \Omega t},
\nonumber\end{eqnarray}
with an arbitrary coefficient $a$ and $J'_m(z)=dJ_m(z)/dz$.
The above expressions hold for real $k_l$, but analytic continuation,  
$k_l\rightarrow i \kappa_l$ with $\kappa_l=\sqrt{q^2-\Omega^2/c_l^2}$, produces the corresponding results
for $\Omega < c_l |q|$. For $R\to \infty$, this step
does not yield physical solutions since $J_m(k_l r)\rightarrow e^{im\pi/2}I_m(\kappa_l r)$ 
diverges for $r\rightarrow \infty$. (The other modified Bessel function, $K_m$, diverges at the origin and is also not acceptable.)  However, such solutions are admitted for finite $R$.  With the replacement $k_l\to k_t$, see Eq.~\eqref{klkt}, the same remarks apply for ${\bf u}_t$ in Eq.~\eqref{utfinal} below.

Next we address the \emph{transverse part}, ${\bf u}_t=\nabla\times{\bm \Psi}$, where
$J_z^{\rm ph}$ acts like (the spin-1 operator $\Sigma_z$ is here expressed in Cartesian coordinates)
\begin{equation}
J_z^{\rm ph} = -i\partial_\phi + \Sigma_z,\quad \Sigma_z = \left(
\begin{array}{ccc}
0 &  -i & 0 \\
i & 0 & 0 \\
0 & 0 & 1
\end{array}
\right).
\end{equation}
The $\Sigma_z$-eigenstates, $\Sigma_z|s\rangle=s|s\rangle$, for the respective eigenvalues ($s=-1,0,1$) are given by 
\begin{equation}\label{spinor}
|1\rangle = \left(\begin{array}{c}
1\\i\\0
\end{array}\right),  \quad 
|0\rangle = \left(\begin{array}{c}
0\\0\\1
\end{array}\right),\quad |-1\rangle = \left(\begin{array}{c}
1\\-i\\0
\end{array}\right).
\end{equation}
In cylindrical coordinates, solutions to Eq.~\eqref{wave2} then have the form 
\begin{eqnarray}
{\bf \Psi}({\bf r},t)&=&  \Bigl( [f_{-1}(r)+f_{1}(r)]\hat e_r +i [f_{-1}(r)-f_{1}(r)]\hat e_\phi \nonumber \\
 &+ &f_0(r)\hat e_z \Bigr)  \frac{e^{im\phi}}{\sqrt{2\pi}}
\frac{e^{iqz}}{\sqrt{L}} e^{-i\Omega t},
\end{eqnarray}
where $f_{s}(r)$ is the radial function for the respective $\Sigma_z$-eigenstate. 
Using $k_t$ in Eq.~\eqref{klkt}, the wave equation \eqref{wave2} then yields Bessel equations that are solved by  
\begin{equation}
    f_{s=-1,0,1}(r) = B_s J_{m+s}(k_t r),
\end{equation} 
with arbitrary constants $B_{s}$.
As a result, we obtain the transverse part of the displacement field as
\begin{eqnarray}\nonumber
{\bf u}_t({\bf r},t) &= &\Biggl[ 
q \left( b_1\frac{m}{k_t r} J_{m}(k_t r) +b_2 J'_{m}(k_t r) \right) \hat e_r \\ \nonumber
&+&  i q \left( b_1J_m'(k_t r) +b_2 \frac{m}{k_t r} J_{m}(k_t r)  \right) \hat e_\phi \\ \label{utfinal}
&-& i k_t b_2 J_m(k_t r) \hat  e_z \Biggr] 
\frac{e^{im\phi}}{\sqrt{2\pi}} \frac{e^{i q z}}{\sqrt{L}}
 e^{-i \Omega t}.
\end{eqnarray}
Due to the constraint $\nabla \cdot{\bf u}_t=0$, here only two 
linear combinations of the three $B_s$ parameters appear, namely $b_1=B_{-1}-B_{1}+ik_t B_0/q$ and $b_2=B_{-1}+B_{1}$.

For a given set $\Lambda$ of conserved phonon quantum numbers (see below),
the normal modes of the displacement field, ${\bf u}_\Lambda({\bf r},t)={\bf u}_l+{\bf u}_t$,
then follow from Eqs.~\eqref{ulfinal} and \eqref{utfinal}.
This result still depends on three arbitrary constants ($a,b_1,b_2$) which
must be determined by geometry-specific boundary conditions and by overall normalization.

\subsection{Cylindrical nanowire}

To calculate the acoustic phonon eigenmodes of an infinitely long cylindrical wire with radius $R$, we now impose stress-free boundary conditions at the surface $r=R$.
After expressing the stress tensor \eqref{stress} in cylindrical coordinates \cite{Landau7}, one finds that 
\begin{equation}
{\bf u}_\Lambda({\bf r},t) = u_r\hat e_r+u_\phi\hat e_\phi+u_z\hat e_z
\end{equation}
has to obey the following boundary conditions at $r=R$,
\begin{eqnarray}
 \label{bc1}
&& iq u_r + \partial_r u_z =0, \quad \partial_r u_\phi  -\frac{u_\phi}{r} + \frac{im}{r} u_r   =0, \\ \nonumber
&& (1-2\xi^2) \left( \partial_ru_r +\frac{u_r}{r} + \frac{im}{r}u_\phi + iqu_z\right) + 2\xi^2 \partial_ru_r=0,
\end{eqnarray}
with $\xi$ in Eq.~\eqref{xidef}.  
We start by solving the simplest case with $m=0$.

\subsubsection{Angular momentum $m=0$}

We first  consider \emph{torsional modes} \cite{Love}, where $u_r=u_z=0$ and only $u_\phi\neq 0$. This corresponds to  the case $a=b_2=0$ in our general solution, where $u_\phi \sim  J_1(k_t r)$.
For $m=0$, the boundary conditions \eqref{bc1} simplify to
$\partial_r u_\phi -u_\phi/r=0$. Inserting our solution, we arrive at a radial quantization condition, 
$J_2(k_tR) =0$, such that only certain eigenfrequencies $\Omega=\Omega_{{\rm T},i}(q)$ (with $i=0,1,\ldots$) are allowed. We obtain
\begin{equation}
\Omega_{{\rm T},i}(q) =c_t \sqrt{q^2+ z^2_{2,i}/R^2},
\end{equation}
where $z_{k,i}$ denotes the non-negative zeroes of the Bessel function $J_k(z)$. The only gapless torsional mode comes from $i=0$ since $z_{2,0}=0$, where we find 
\begin{equation}\label{tors0}
\Omega_{{\rm T}}(q)=c_t|q|,\quad {\bf u}^{\rm T}_ q({\bf r}) = \frac{2r}{R^2} \frac{e^{iqz}}{\sqrt{2\pi L}} \hat e_\phi. 
\end{equation} 
For all $i>0$, the dispersion relation acquires the finite gap $z_{2,i} c_t/R$. Using $z_{2,1}\simeq 5.1356$, we estimate the smallest of these gaps as $\approx 34$~meV for Bi$_2$Te$_3$ wires of radius $R\approx 100$~nm. 
Staying on energy scales well below this gap, we can neglect all gapped torsional phonon modes.  This step is assumed in our low-energy theory from now on where only the $i=0$ torsional mode in Eq.~\eqref{tors0} will be retained.
We note that torsional modes cannot exist for $\Omega<c_t|q|$, 
since the modified Bessel function $I_2(\kappa_t R)$ obtained after analytic continuation has no zeroes except at the origin.

All other phonon eigenmodes for $m=0$ follow by setting
$u_\phi =0$, corresponding to $b_1=0$ in our general expression for ${\bf u}_\Lambda$.
The boundary conditions \eqref{bc1} then yield the condition 
${\bm M} (a,b_2)^T=0$, with the matrix ${\bm M}$ given by 
\begin{equation}\label{mdef0}
\left(\begin{array}{cc}
qk_l J_1(k_lR) &  -(k_t^2-q^2) J_1(k_tR)\\
(k_t^2-q^2) J_0(k_lR) - \frac{2k_lJ_1(k_lR)}{R} & 
4qk_tJ_1'(k_tR)\end{array}
\right).
\end{equation}
A non-trivial solution exists only for ${\rm det}{\bm M}=0$, which yields the 
radial quantization condition in the form of Pochhammer's frequency equation \cite{Graff},
\begin{eqnarray}
&& 4q^2k_l k_t J_1(k_lR) J_0(k_tR) +(k_t^2-q^2)^2 J_1(k_tR)J_0(k_lR) \nonumber \\
&& \qquad \qquad  = \, \frac{2k_l \Omega^2}{Rc_t^2}J_1(k_lR) J_1(k_tR).
\label{disprelm=0}
\end{eqnarray}
As we show next, Eq.~\eqref{disprelm=0} describes both 
longitudinal modes \cite{Love} for $\Omega>c_t |q|$, and Rayleigh surface modes for $\Omega<c_t|q|$.

We start with the \emph{longitudinal phonon modes}.
First, for $q \to 0$, Eq.~\eqref{disprelm=0} simplifies [with $\xi$ in Eq.~\eqref{xidef}] to 
\begin{equation}\label{disprelm=0q=0}
  J_1(\varpi) \left[\varpi J_0(\xi \varpi) - 2\xi J_1(\xi \varpi)\right]=0, \quad \varpi\equiv R\Omega/c_t.
\end{equation}
 Noting that $\varpi=\Omega=0$  solves Eq.~\eqref{disprelm=0q=0}, we observe that
  a gapless longitudinal phonon mode will always exist.
In addition, Eq.~\eqref{disprelm=0q=0} admits  gapped longitudinal modes as for the torsional case, which we again do not take into account  in  the low-energy theory. 
Second, for long wavelengths, $|q|R\ll 1$, by expanding the Bessel functions
in Eq.~\eqref{disprelm=0}, we find the dispersion relation for the gapless longitudinal mode \cite{Love},
\begin{equation}\label{longitudinal-smallR}
\Omega_{\rm L}(q) = c_{\rm L} |q| \left[ 1 -(\sigma q R/2)^2 \right] + {\cal O} \left( |qR|^5 \right) . 
\end{equation}
The sound velocity for this mode is given by $c_{\rm L}=\sqrt{E/\rho_M}$ with the Young modulus $E=2(1+\sigma)\mu$. For Bi$_2$Te$_3$, the value for $E$ in Refs.~\cite{Huang2008,Jenkins1972} results in $c_{\rm L}\simeq 2500$~m$/$s.
In Eq.~\eqref{longitudinal-smallR}, we also use Poisson's ratio $\sigma=\lambda/ [2(\lambda+\mu)]$. Since usually the latter quantity is within the window $0< \sigma< 1/2$, we find that 
$c_t < c_{\rm L} < c_l$.  As a consequence, 
 the longitudinal mode \eqref{longitudinal-smallR} has imaginary wave number $k_l=i\kappa_l$ but  real wave number $k_t$. 

At short wavelengths,  $|q|R\gg 1$, the longitudinal mode evolves into a \emph{Rayleigh  mode} with $\Omega< c_t|q|$. From Eq.~\eqref{disprelm=0}, after
analytic continuation  $k_{l,t} \rightarrow i\kappa_{l,t}$,  no physical solutions are found for $|q|R\ll 1$, i.e., there are no cylindrical Rayleigh waves in the long wavelength limit.
However, for $|q|R \gg 1$, asymptotic expansion of Eq.~\eqref{disprelm=0} shows that Rayleigh modes do exist at short wavelength, with dispersion relation
\begin{equation}  \label{Rayleighmode}
\Omega_{\rm R}(q) = c_{\rm R}| q |+ \frac{\eta_0 c_{\rm R}}{ R} + {\cal O}\left( \frac{1}{|q|R} \right),
\end{equation}
where $c_{\rm R}=\zeta c_t$ is the Rayleigh mode velocity for a planar surface \cite{Giraud2011,Sirenko1997} which follows by letting $R\to \infty$.
The dimensionless number $\zeta<1$ is a lengthy function of $\xi=c_t/c_l$, 
 with $\zeta\simeq 0.92$ for Bi$_2$Te$_3$ \cite{Giraud2011}. 
In Eq.~\eqref{Rayleighmode}, we also use the number
\begin{equation}
 \eta_0 =\frac{\gamma_t(1-\gamma_l\gamma_t)}{2\zeta^2[2\sqrt{\gamma_l\gamma_t }-\xi^2\gamma_t/\gamma_l- \gamma_l/\gamma_t]},
\end{equation}
with $\gamma_{t}=\sqrt{1-\zeta^2}$ and $\gamma_l=\sqrt{1-\zeta^2\xi^2}$. 

The longitudinal mode with $\Omega_{\rm L}(q)\simeq c_{\rm L}|q|$ thus gradually evolves into the Rayleigh mode 
with $\Omega_{\rm R}(q)\simeq c_R |q|$ as $|q|R$ increases.
Since we here focus on the low-energy regime, only the longitudinal mode will be kept in what follows.
To leading order in $|q|R\ll 1$, we  find the dispersion relation and the normalized eigenmode as
\begin{equation}\label{long0}
    \Omega_{\rm L}(q) = c_{\rm L} |q| ,\quad {\bf u}^{\rm L}_q({\bf r}) =  \frac{\sqrt2 \, {\rm sgn}(q) }{R}\frac{e^{iqz}}{\sqrt{2\pi L}} \left( \sigma q r \hat e_r+i\hat e_z\right).
\end{equation}

\subsubsection{Angular momentum $m\ne 0$}

We now briefly turn to the  case of finite phonon angular momentum, $m\neq 0$.
The boundary conditions \eqref{bc1} then yield the condition
${\bm M}_m(a,b_2,b_1)^T  = 0$, where the $m=0$ matrix ${\bm M}$ in Eq.~\eqref{mdef0}
is replaced by 
\begin{widetext}
\begin{equation}
{\bm M}_m = \left( \begin{array}{ccc}
q k_l J_m^{l \prime}    & (q^2-k_t^2) J_m^{t\prime} &  \frac{mq^2}{k_tR} J_m^{t} \\
 (q^2+\frac{2m^2}{R^2} -k_t^2) J_m^l - \frac{2k_l}{R} J_m^{l\prime} &
qk_t(J^t_{m-2}+J^t_{m+2} -2 J_m^t)  & qk_t (J^t_{m-2}-J^t_{m+2}) \\
\frac12 k_l^2 (J_{m-2}^l-J_{m+2}^l) &  qk_t(J_{m-2}^t - J_{m+2}^t) & qk_t(J_{m-2}^t+J_{m+2}^t) 
\end{array} \right).
\end{equation}
\end{widetext}
We use the shorthand
notations $J_m^{l,t}\equiv J_m(k_{l,t}R)$ and 
$J_m^{l,t}{}^\prime$  for the respective derivative.
One easily checks that for $m=0$, the above results are recovered from these expressions.

For $|q|R\ll 1$ and $m=\pm 1$, one obtains flexural modes with a quadratic dispersion relation \cite{Love},
\begin{equation}
\Omega_{\rm F}(q)= c_{\rm L}R q^2 + {\cal O}\left(|qR|^3\right).
\end{equation}
These are the energetically lowest phonon modes in a cylindrical wire at long wavelengths.  
However, for the deformation potential coupling in Sec.~\ref{sec4}, we will find that only $m=0$ phonons couple to electrons.  For that reason, we do not discuss $m\neq 0$ 
phonon modes in more detail here.

\subsection{Quantization}

The quantization of the phonon theory now proceeds along standard paths.  The displacement field is expressed in terms of bosonic annihilation operators, $b^{}_\Lambda$, with the commutation relation
$[ b_\Lambda^{\phantom{\dagger}}, b^{\dagger}_{\Lambda'} ] =  \delta_{qq'} \delta_{mm'}\delta_{\lambda\lambda'},$
where $\Lambda$ denotes the set of quantum numbers $(q,m,\lambda)$. The index $\lambda$ labels the different branches (e.g., torsional or longitudinal modes) and, in general, includes gapless as well as gapped modes.  Inserting the eigenfrequencies $\Omega=\Omega_\Lambda$ into the above normal mode expressions $u_\Lambda({\bf r},t)$, 
we have 
\begin{equation}\label{displfinal}
{\bf u}({\bf r},t) = \sum_\Lambda \frac{1}{\sqrt{2\rho_M L \Omega_\Lambda}} {\bf u}_\Lambda({\bf r},t)  b_\Lambda + {\rm h.c.},
\end{equation}
with the non-interacting second-quantized phonon Hamiltonian  
\begin{equation} \label{phHam}
H_{\rm ph} = \sum_\Lambda \Omega_\Lambda\left( b^\dagger_\Lambda b^{\phantom{\dagger}}_\Lambda + 1/2 \right).
\end{equation}

We end this section by briefly summarizing the above results as far as we need them in what follows. 
As shown in the next section, the only gapless phonon branch that couples to electrons via the deformation potential is given by longitudinal 
phonons with zero angular momentum.  Their dispersion relation and the corresponding normal mode expression
are specified for $|q|R\ll 1$
in Eq.~\eqref{long0}.  All other phonon branches are either gapped (and can thus be included by a renormalization of the Luttinger
liquid parameter), or they do not couple to electrons within our low-energy theory.

\section{Electron-phonon coupling} \label{sec4}

\subsection{Deformation potential}

We next turn to the electron-phonon coupling, assuming that the dominant contribution arises from the deformation potential, cf.~Refs.~\cite{Giraud2011,Giraud2012},
\begin{equation}\label{defpot}
    H_{\rm e-ph}= \alpha  \int d{\bf r} \rho_{\rm el}({\bf r}) {\rm Tr} u ({\bf r}),
\end{equation}
where Ref.~\cite{Huang2008} estimates the bare coupling strength $\alpha\approx 35$~eV for Bi$_2$Te$_3$.
However, this value  could be significantly reduced by internal screening effects and we only use it as a rough estimate.
Inserting Eq.~\eqref{densop} for the electronic density operator, we first notice 
that since we keep only a single Dirac fermion subband corresponding to TI surface states with angular momentum $j=-\Phi_B$, 
 only phonon modes with angular momentum $m=0$ can couple to electrons.  
 (This statement continues to be valid in Sec.~\ref{sec6b}, where we study slight deviations of $\Phi_B$ from half-integer values. 
 Also in such a case, only a single subband needs to be kept.)
 Moreover,
at low energy scales, we can restrict ourselves to the gapless torsional and longitudinal phonon modes with $m=0$, see Eqs.~\eqref{tors0} and \eqref{long0}, respectively.  
Since ${\rm Tr}u=\nabla\cdot {\bf u}^{\rm T}=0$ for the torsional phonon mode in Eq.~\eqref{tors0}, 
the only contribution of the deformation potential \eqref{defpot} to our low-energy theory arises from the $m=0$ longitudinal phonon mode in Eq.~\eqref{long0}.
Assuming that only phonon momenta with $|q|R\ll 1$ are important and
taking the continuum limit $L\to\infty$ in Eqs.~\eqref{displfinal} and \eqref{phHam}, we find 
\begin{eqnarray}\nonumber
\nabla\cdot {\bf u}({\bf r},t)&=& -(1-2\sigma) \int \frac{dq}{2\pi} \frac{|q|}{\sqrt{2\bar\rho\Omega_q}}e^{iqz} \times \\ &\times&
\left(b_q^{} e^{-i\Omega_qt} +b_{-q}^\dagger e^{i\Omega_qt}\right),
\end{eqnarray}
and
\begin{equation}
 H_{\rm ph} = \int\frac{dq}{2\pi} \ \Omega_q \left( b^\dagger_q b_q^{} + 1/2 \right), \quad \Omega_q\equiv c_{\rm L}|q|. \label{hph2}
\end{equation}
Here the linear mass density is given by $\bar\rho=\pi R^2 \rho_M$,
and the phonon operators $b_q$ refer to $m=0$ longitudinal modes, with commutator $[b_q^{},b_{q'}^\dagger]=2\pi\delta(q-q')$.

Using Eqs.~\eqref{densop}, \eqref{long0}, and \eqref{displfinal}, we then find from Eq.~\eqref{defpot}  
the coupling Hamiltonian
\begin{equation}\label{He-ph}
    H_{\rm e-ph}= - \frac{i\alpha(1-2\sigma)}{c_{\rm L} }  \int \frac{dq}{2\pi}  \sqrt{\frac{\Omega_q}{2\pi\bar \rho}} \ q\tilde\theta(q) \left( b_q^{}+b_{-q}^{\dagger}\right) ,
\end{equation}
where $\theta(z)= \int \frac{dq}{2\pi} e^{iqz}\tilde\theta(q)$ with $[\tilde \theta(q)]^\dagger=\tilde\theta(-q)$ is the boson field introduced in Sec.~\ref{sec2}.

For half-integer flux $\Phi_B$, the coupled electron-phonon problem can now be solved exactly even in the presence
of electron-electron interactions. 
We proceed in analogy to~Refs.~\cite{Loss1992,DeMartino2003}, where non-helical Luttinger liquids coupled to acoustic phonons have been studied.
The Euclidean action for the entire system, $S=S_{\rm el}+S_{\rm ph}+S_{\rm e-ph}$, follows from the low-energy Hamiltonian terms in Eqs.~\eqref{helec}, \eqref{hph2} and \eqref{He-ph}, respectively.
Instead of the $b_q$ and $b_q^\dagger$ phonon operators, it is convenient to use the oscillator amplitude operators 
\begin{equation}
    u_q= \frac{1}{\sqrt{2\Omega_q}} \left( b_q^{}+b_{-q}^\dagger \right),\quad p_q=-i\sqrt{\frac{\Omega_q}{2}} \left( b_q^{}-b_{-q}^\dagger \right),
\end{equation}
with the commutator $[u_q,p_{q'}]=2\pi\delta(q+q')$.  Using bosonic Matsubara frequencies, $\omega_n=2\pi nT$ (integer $n$), the dependence on imaginary time $\tau$ is 
resolved by the expansion
\begin{equation}
u_q(\tau)= T\sum_{\omega_n} e^{-i\omega_n\tau} \tilde u_q(\omega_n),\quad \tilde u_q^\ast(\omega_n)= \tilde u_{-q}^{}(-\omega_n),
\end{equation}
and likewise for $p_q(\tau)$ and $\theta_q(\tau)$.  
With the shorthand notation 
\begin{equation}
    \int [dq]\ (\cdots)= T\sum_{\omega_n}\int \frac{dq}{2\pi} \ (\cdots),
\end{equation}
and writing $\tilde u_q(\omega)\to u_q(\omega)$ 
(and so on), we obtain the action contributions 
\begin{eqnarray}\nonumber
    S_{\rm el}&=& \frac{1}{2vK}\int [dq]\left(\omega_n^2+v^2 q^2\right) \left| \theta_q(\omega_n)\right|^2,\\ \label{actionfull}
    S_{\rm ph}&=& \frac12 \int [dq] \left(\omega_n^2+\Omega_q^2\right) \left| u_q(\omega_n)\right|^2,\\ \nonumber
S_{\rm e-ph}&=& \frac{i\alpha(1-2\sigma)}{\pi\sqrt{\bar\rho}} \int [dq] \ {\rm sgn}(q) q^2 u_q(\omega_n) \theta_{-q}(-\omega_n). 
\end{eqnarray}
Here $v\equiv v_1/K$ is the plasmon velocity in the helical Luttinger liquid. We recall that $v_1$ is the Fermi velocity along the wire axis and $K$ the Luttinger liquid parameter. In practice, one has $v\gg c_{\rm L}$. 
We thus arrive at an exactly solvable Gaussian functional integral for the coupled electron-phonon system.

\subsection{Integrating over the phonon degrees of freedom}

In this work, we focus on the electronic degrees of freedom and therefore proceed by integrating over the phonon amplitudes $u_q(\omega_n)$. 
As a result of this Gaussian functional integration, the effective action for the bosonized $\theta$ field describing the electronic sector is given by
\begin{equation}\label{effact}
    S_{\rm eff}= \frac12 \int[dq] \ D_{\theta\theta}^{-1}(\omega_n,q)\left |\theta_q(\omega_n)\right|^2,
\end{equation}
with the inverse propagator
\begin{equation}\label{dthth}
    D^{-1}_{\theta\theta}(\omega_n,q) = \frac{1}{vK} \left( \omega^2_n + v^2 q^2 - \left(\frac{AvK}{c_{\rm L}}\right)^2\frac{\Omega_q^4}{\omega_n^2+\Omega_q^2} \right),
\end{equation}
where we define the dimensionless electron-phonon coupling parameter 
\begin{equation}\label{bdef}
    A = \frac{(1-2\sigma)\alpha}{\pi c_{\rm L} \sqrt{\bar \rho v_1}}. 
\end{equation}
Inserting theoretical estimates for the parameters in Eq.~\eqref{bdef}  for Bi$_2$Se$_3$ and/or Bi$_2$Te$_3$ \cite{Huang2008,Jenkins1972}, 
one finds typical values of the order $A\alt 1$. Our approach represents a controlled approximation in the low-energy regime.  
In particular, we assume that the relevant energy scales are well below $v_1/R$ such that higher electronic subbands can be neglected.  
However, gapped phonon bands could be included by a renormalization of the Luttinger liquid parameter \cite{Gogolinbook}, and 
we only have to explicitly retain the gapless phonon mode considered above.

With the velocities $v_\pm>0$ defined from 
\begin{equation}\label{vsdef}
    v_\pm^2 = \frac12\left( v^2+c_{\rm L}^2 \pm \sqrt{(v^2-c_{\rm L}^2)^2+ (2A vK c_{\rm L})^2} \right)
\end{equation}
and the  residues  
\begin{equation}\label{fsdef}
    F_\pm = \frac{v_\pm^2-c_{\rm L}^2}{v_\pm^2-v_\mp^2},
\end{equation}
the propagator follows as
\begin{equation}
    D_{\theta\theta}(\omega_n,q) =vK \sum_{s=\pm} \frac{F_s}{\omega_n^2+v_s^2q^2}.
\end{equation}
We note that Eq.~\eqref{fsdef} implies $F_++ F_-=1$ and $\sum_s F_s (v/v_s)^2= 1/(1-A^2K^2)$.
Similarly, the propagator for the dual boson field $\varphi$ in Eq.~\eqref{1Dop} follows as
\begin{equation}\label{gsdef}
      D_{\varphi\varphi}(\omega_n,q) = \frac{1}{vK}\sum_{s=\pm} \frac{v_s^2 F_s}{\omega_n^2+v_s^2q^2},
\end{equation}
For $A=0$, one finds $v_+=v$ and $v_-=c_{\rm L}$, with $F_+=1$ and $F_-=0$. 

Using the above expressions, the electronic Green's function,
\begin{equation}
    {\cal G}({\bf r},\tau)=-\left\langle{\cal T}_\tau \Psi^{}_{\rm el}({\bf r},\tau)\Psi^\dagger_{\rm el}(0,0)\right\rangle,
\end{equation}
with the electron operator in Eq.~\eqref{1dexp} and the time ordering operator ${\cal T}_\tau$, can  be computed in an exact manner.   The  nontrivial $(z,\tau)$ dependence,
${\cal G}(z,\tau)\propto \sum_{\nu=\pm} e^{i\nu k_Fz} G_\nu(z,\tau)$, follows from the 1D Green's functions, 
\begin{equation}
    G_\nu(z,\tau)=-\langle{\cal T}_\tau \psi^{}_\nu(z,\tau)\psi^\dagger_\nu(0,0)\rangle,
\end{equation}
where off-diagonal contributions (with $\nu\ne \nu'$)  vanish identically.  
Using the bosonized 1D operators in Eq.~\eqref{1Dop}, we obtain 
\begin{eqnarray}\nonumber
    G_{\nu=\pm}(z,\tau)&=& \frac{{\rm sgn}(\tau)}{4\pi\xi_\perp} \prod_{s=\pm} \left| \frac{\xi_\perp}{z+iv_s \tau}\right|^{\left(\frac{vK}{2v_s}+\frac{v_s}{2Kv}\right)F_s}  \\
    &\times& \left(\frac{z+i\nu v_s\tau}{z-i\nu v_s\tau} \right)^{F_s}.
\end{eqnarray}
Given this result, one can compute the spectral function from the imaginary part of ${\cal G}$, cf.~Refs.~\cite{Gogolinbook,Schulz2010}.  
The latter quantity could in principle be measured by photoemission spectroscopy.  However, in what follows we shall focus on simpler observables.

\section{Phase diagram}\label{sec5}

We next turn to the zero-temperature phase diagram of the coupled electron-phonon system with half-integer flux $\Phi_B$.  The effective low-energy action \eqref{effact} for the electronic sector, obtained after integration over the phonon degrees of freedom, allows us to obtain the  exact correlation functions of all possible order parameters.  In this 1D system, long-range order is not possible and one can at best find an algebraic decay of correlation functions (at $T=0$). It is  then common  practice to define the phases according to the smallest decay exponent \cite{Gogolinbook}.   
For extremely strong electron-phonon couplings with $A\ge 1/K$ in Eq.~\eqref{bdef},
one encounters the so-called Wentzel-Bardeen singularity, where the system becomes unstable and undergoes phase separation \cite{Loss1992,DeMartino2003}.  
In what follows, we assume that $A<1$ and the system is stable.
We then examine different candidate order parameter correlations.  

First, as pointed out in Sec.~\ref{sec2}, charge density wave (CDW) correlations cannot exist in our system due to the absence of $2k_F$ backscattering.  However, spin density wave (SDW) correlations are possible.
For the surface state of the TI wire, we can either have a spin density operator component $s_\phi$ along the circumferential direction, or a component $s_z$ along the wire axis.  In bosonized form, they are given by 
\cite{Egger2010}
\begin{eqnarray}
    s_\phi(z,\tau) &= &\frac{1}{2\sqrt{\pi}} \partial_z\varphi(z,\tau),\\ \nonumber
    s_z(z,\tau) &=&- \frac{1}{2\pi\xi_\perp}  \cos[2k_Fz+2\sqrt{\pi} \theta(z,\tau)].
\end{eqnarray}
The first relation is due to spin-momentum locking of the TI surface state: the current density operator along the $z$-axis has precisely the same form. 
We obtain the $T=0$ correlation functions (the mixed correlator vanishes) 
\begin{eqnarray}\nonumber
    \left\langle s_z(z,\tau) s_z(0,0)\right\rangle &\propto & \cos(2k_Fz) \prod_{s=\pm} \left|\frac{\xi_\perp}{z+iv_s\tau}\right|^{2vKF_s/v_s},\\ 
    \left\langle s_\phi(z,\tau) s_\phi(0,0)\right\rangle &\propto & \prod_s \left| z+iv_s\tau\right|^{-\nu_\phi/2},
\end{eqnarray}
which yields the corresponding decay exponents  $\nu_z=2vK\sum_s F_s/v_s$ and $\nu_\phi=2$. Here the $F_\pm$ have been defined in Eq.~\eqref{fsdef}.  
For material parameters where $\nu_\phi$ represents the slowest decay, we call the phase `metallic' since here the current-current correlations
have the same decay law as in an unperturbed Luttinger liquid. 
Next, the order parameter for singlet superconductivity is proportional to 
${\cal O}_{\rm sc}(z,\tau)= \psi_+(z,\tau) \psi_-(z,\tau)\propto e^{2i\sqrt{\pi}\varphi}$ \cite{footnew,Lutchyn2019a}. 
Pairing correlations thus decay along the wire direction as
\begin{equation}
    \langle {\cal O}_{\rm sc}(z,\tau){\cal O}_{\rm sc}^\dagger(0,0)\rangle \propto \prod_{s=\pm} \left| \frac{\xi_\perp}{z+iv_s\tau}\right|^{2v_s F_s/(Kv)}.
\end{equation}
The resulting decay exponent is given by $\nu_{\rm sc}=(2/vK)\sum_s v_sF_s$.

Using the above results for the three exponents ($\nu_z,\nu_\phi,\nu_{\rm sc}$), 
the phase diagram  in the $K$-$A$ plane is readily determined by finding the smallest exponent for given parameter choice, 
see Fig.~\ref{fig1} in Sec.~\ref{sec1}.
For our TI nanowire pierced by a half-integer flux $\Phi_B$, the radius $R$ appears only implicitly via the definition
of the dimensionless electron-phonon coupling parameter $A$ in Eq.~\eqref{bdef}, and possibly through a weak $R$-dependence of the Luttinger liquid parameter $K$ \cite{Egger2010}. 
The latter parameter can encode both the effects of Coulomb interactions and those of residual optical phonon modes not taken into account in our model, cf.~Ref.~\cite{Gogolinbook}, where
$K=1$ in the absence of interactions, $K<1$ for repulsive interactions, and $K>1$ for effectively attractive interactions.
In Fig.~\ref{fig1}, we show the phase diagram using parameters appropriate for the TI material HgTe, with $v_1\simeq 5\times 10^5$~m$/$s \cite{Ziegler2018} and $c_{\rm L}\simeq 2400$~m/s 
\cite{Kurilo1997}. 
The HgTe case is especially interesting since it has been established by recent nanowire experiments 
that the ballistic regime is reachable in practice \cite{Ziegler2018}.  We thus expect that our predictions can be tested in the immediate future.
We note that the phase diagram for Bi$_2$Te$_3$ looks qualitatively very similar.

In the absence of electron-electron interactions ($K=1$), we observe that superconducting correlations dominate for arbitrary electron-phonon coupling strength $0<A<1$, in accordance with earlier studies for non-helical Luttinger liquids \cite{Loss1992,DeMartino2003}.  
Unless electron-electron interactions are screened off, however, we expect that the superconducting correlations are quickly overcome by SDW correlations which are favored for $K<1$
and small values of $A$.  For large $A$ (but $A<1/K$), we also find a tiny intermediate metallic phase, see inset of Fig.~\ref{fig1}. 
The phase boundary curves separating the metallic phase from the SDW and the SC phases, respectively, can be analytically 
shown to merge at the special point $(K=1,A=0)$.  However, no merging point exists in the limit $K\to 0$.
Ultimately, for $A\ge 1/K$, the system becomes unstable.  

Our theory therefore suggests the possibility of dominant intrinsic pairing fluctuations when Coulomb interactions are well screened off. 
The resulting superconducting wire could then even harbor Majorana bound states, see Ref.~\cite{Cook2012}. 
Such states can exist even in 1D wires with intrinsic superconducting pairing \cite{Fidkowski2011}.
However,  we expect that proximity-induced superconductivity will be needed in practice to achieve this goal since the relevant energy 
scales protecting the Majorana states will otherwise be tiny.

Finally, we note that the phase diagram can significantly change when $\Phi_B$ does not have half-integer values.  
As we discuss in detail in Sec.~\ref{sec6b}, the presence of $2k_F$ scattering then implies that also regions with CDW ordering 
become possible.  We leave the exploration of the phase diagram for general $\Phi_B$ to future work.

\section{Phonon-induced resistivity} \label{sec6}

We now turn to the phonon-induced electrical resistivity, $\rho$, of a long cylindrical TI nanowire pierced by a magnetic flux $\Phi_B$, taking into account electron-phonon couplings of dimensionless strength $A<1/K$, see Eq.~\eqref{bdef}.  
We start in Sec.~\ref{sec6a} with the case of half-integer flux $\Phi_B$ for arbitrary Luttinger liquid parameter $K$. From the Kubo formalism, we show that phonons do not generate a finite resistivity correction $\rho(T)$ due to the absence of $2k_F$-backscattering processes.   In Sec.~\ref{sec6b}, focusing on the case without electron-electron interactions ($K=1$), we  
allow for small flux deviations $\delta \Phi_B$  away from half-integer values.   Backscattering then becomes possible and one obtains a finite resistivity for $T>0$.  
For quantitative results, we follow the  
Boltzmann equation approach of Ref.~\cite{Gurevich1995}.  Alternatively, one could proceed along the bosonization route of Ref.~\cite{Seelig2005}, which 
also allows to cover the $K\ne 1$ case for $\delta \Phi_B\ne 0$.  However, in Sec.~\ref{sec6b} we confine ourselves to the physically transparent Boltzmann approach for $K=1$.

\subsection{Half-integer flux: Kubo formula}\label{sec6a}

We begin with the case of precisely half-integer flux $\Phi_B$ and start from the Kubo formula for the $(\omega,q)$-dependent conductivity \cite{Gogolinbook},
\begin{equation}\label{cond1}
    \sigma(\omega,q)= \frac{i}{\omega} \left( \frac{e^2 v K}{\pi} + \Pi(\omega,q)\right),
\end{equation}
where $\Pi(\omega,q)$ is the retarded current-current correlation function. The latter quantity is first computed in Matsubara frequency space, 
\begin{equation}
    \Pi(i\omega_n, q) = - \left\langle J^*(i\omega_n,q) J(i\omega_n,q) \right\rangle_{S_{\rm eff}},
\end{equation}
followed by the analytic continuation $i\omega_n\to \omega + i 0^+$.
The charge current operator is here given by $J=\frac{evK}{\sqrt{\pi}} \partial_z\varphi$ \cite{Gogolinbook}.  
Using Eq.~\eqref{gsdef}, we obtain 
\begin{equation}
    \Pi(i\omega_n,q) = \frac{e^2 v K}{\pi} \left( -1 + \sum_{s=\pm}  \frac{\omega_n^2}{\omega_n^2+v_s^2q^2} F_s \right).
\end{equation}
Performing the analytic continuation, Eq.~\eqref{cond1} yields 
\begin{eqnarray}\nonumber
    \sigma(\omega,q) &=& \frac{e^2 v K }{2\pi} \sum_{s=\pm,\nu=\pm} F_s \\ \nonumber
    & \times & \left( \pi \delta(\omega-\nu v_s q) + i{\cal P}\frac{1}{\omega-\nu v_s q}\right),\label{sss} 
\end{eqnarray}
where ${\cal P}$ denotes the principal part and the velocities $v_\pm$ have been specified in Eq.~\eqref{vsdef}. 
We thus obtain
\begin{equation}
\lim_{q\to 0} {\rm Re}\sigma(\omega,q) = e^2 v K \delta(\omega) \sum_{s=\pm} F_s= e^2 vK \delta(\omega).  
\end{equation}
The real part of the conductivity yields a $\delta$-function Drude peak at $\omega=0$ for $q\to 0$, and hence a vanishing resistivity at all temperatures (where the above model applies). 
Since $vK=v_1$ by Galilean invariance, neither electron-electron nor electron-phonon interactions cause corrections to the conductivity.
This result is rationalized by the absence of $2k_F$-backscattering processes in TI nanowires pierced by a precisely half-integer flux $\Phi_B$.   
In the next subsection, we address what happens when $\Phi_B$ deviates from half-integer values. 

\subsection{Away from half-integer flux}\label{sec6b}

We now focus on the case without electron-electron interactions, $K=1$, and study the effects of a static deviation of $\Phi_B$ from half-integer values, $\delta\Phi_B\ne 0$.
Such a situation may arise either due to changes in the magnetic field strength or its direction, or from fluctuations of the cross-sectional area of the nanowire.
For simplicity, we assume $|\delta \Phi_B|\ll 1$ below. 
For $\delta\Phi_B\ne 0$, since the electron density operator (\ref{densop}) will now have a $2k_F$-oscillatory contribution due to the absence of topological protection, 
phonons can cause electron backscattering. 
 We then expect a temperature-dependent correction to the electrical conductance of a TI nanowire.
To study this effect in quantitative terms, we follow Ref.~\cite{Gurevich1995} and use the Boltzmann equation to evaluate the phonon-induced conductance correction for
a long TI nanowire of length $L$.  Without coupling to phonons ($A=0$), the ballistic system has the quantized and temperature-independent 
conductance $G=G_0=e^2/h$ \cite{Egger2010}. 

To determine the low-energy form of the electron density operator, we first generalize the electron operator in Eq.~\eqref{1dexp} to the case $\delta\Phi_B\ne0$.
At low energies, we may focus on the single band with total angular momentum $j$ such that $\Phi_B=-j+\delta\Phi_B$.  
Assuming that the chemical potential $\mu$ is located in the conductance band, Eq.~\eqref{ener2} implies that the Fermi momentum is now given by
\begin{equation}\label{gammadef}
k_F \simeq \frac{\mu}{v_1} \left( 1 - 2\gamma^2 \right),\quad  \gamma= \frac{v_2 \delta \Phi_B}{2\mu R}.
\end{equation}
We here study the consequences of $\gamma\ne 0$ to leading order in $\gamma$, i.e., for $|\gamma|\ll 1$.
Taking the conduction band eigenstate of ${\cal H}_{\rm el}(k)$ in Eq.~\eqref{helec} with angular momentum $j$ from Ref.~\cite{Egger2010}, the low-energy electron operator 
follows as
\begin{eqnarray}\nonumber
\Psi_{\rm el} (r,\phi,z)&=& \frac{f_\perp(r)}{\sqrt{4\pi}} \sum_{\nu=\pm} e^{i\nu k_F z} e^{i(j-1/2)\phi}\psi_\nu(z) \\ && \qquad \quad \times  \ 
\left(\begin{array}{c} \nu(1-\gamma)  \\  i (1+\gamma)e^{i\phi} \end{array}\right) ,
\end{eqnarray}
where we drop all ${\cal O}(\gamma^2)$ terms.
As for $\gamma=0$ in Eq.~\eqref{1dexp}, the 1D field operators $\psi_{\nu=\pm}(z)$ describe right- or left-moving fermionic quasiparticles. 
Indeed, linearization of the dispersion relation \eqref{ener2} around the respective Fermi point, $k=\nu k_F+p$ with $|p|\ll k_F$, yields
$E_{\nu=\pm}(p) \simeq \mu \pm v_1 p$.

The 1D electron density operator, $\rho_{\rm 1D}(z)$,  is obtained by integration over
the cross section of the nanowire and follows (to leading order in $\gamma$) as
\begin{eqnarray}\nonumber
\rho_{\rm 1D}(z)&=&\int rdr d\phi \, \Psi^\dagger_{\rm el}({\bf r})
\Psi^{}_{\rm el} ({\bf r})\\ \label{2kf}
& =& \sum_{\nu=\pm}  \psi^\dagger_\nu \psi_{\nu}^{}  + 2\gamma \sum_{\nu} e^{-i\nu 2k_Fz} \psi^\dagger_\nu \psi^{}_{-\nu}\\
\nonumber
&=&
\frac{1}{\sqrt{\pi}}\partial_z\theta(z)  +  \frac{2\gamma}{\pi \xi_\perp} \cos \left[2k_F z +2 \sqrt{\pi} \theta(z)\right].
\end{eqnarray}
In the last step, we have used the bosonization identity \eqref{1Dop}.
Equation \eqref{2kf} shows that for $\delta \Phi_B \neq 0$, the electron density operator contains a $2k_F$-oscillatory term 
corresponding to electron backscattering.  By variation of the  flux $\delta \Phi_B$, the relative importance of this term compared to the forward scattering 
contribution --- the first term in Eq.~\eqref{2kf} --- can be changed.   
For $\gamma\ne 0$, on top of Eq.~\eqref{He-ph} the electron-phonon interaction Hamiltonian then receives an additional term from the
deformation potential in Eq.~\eqref{defpot}, 
\begin{eqnarray}\nonumber
H'_{\rm e-ph} &=&- v_1 Z \int dz\sum_{\nu=\pm} e^{-i\nu 2k_Fz} \psi^\dagger_\nu \psi^{}_{-\nu} \\
&\times&\label{newheph}
\int \frac{dq}{2\pi}  e^{iqz} \sqrt{|q|} \left(b^{}_q+b^\dagger_{-q}\right),
\end{eqnarray}
which describes electron backscattering with the simultaneous absorption or emission of a phonon. 
The corresponding dimensionless coupling constant is given by
\begin{equation}\label{Zdef}
Z = \sqrt{2 \pi^2 c_{\rm L}/v_1}\ A\gamma ,
\end{equation}
with the electron-phonon coupling parameter $A$ in Eq.~\eqref{bdef} and  $\gamma\propto \delta\Phi_B$ in Eq.~\eqref{gammadef}.

The transition probability for absorption (`$-$') or emission (`$+$') of a phonon during a quasiparticle scattering process with momentum $p\to p'$ with respect to the Fermi points $\nu\to \nu'$ can be estimated from Fermi's golden rule as  
\begin{equation}
    W^\pm_{\nu',\nu}(p',p)\propto \Omega_{p-p'}\ \delta\left(E_{\nu'}(p')-E_{\nu}(p)\pm \Omega_{p-p'+(\nu-\nu')k_F} \right).
\end{equation}
Using the linearized dispersion relation
$E_\nu(p)=\mu+\nu v_1p$, we first observe that energy conservation requires $v_1 |p-p'| =c_{\rm L} |p-p'|$ for
forward scattering processes ($\nu'=\nu$).
Unless one accidentally has $c_{\rm L}= v_1$, the only solution is given by $p=p'$. Transition probabilities for forward
scattering processes thus vanish identically, $W^\pm_{\nu,\nu}(p',p)=0$, in accordance with our results in Sec.~\ref{sec6a}.
For $\gamma\ne 0$, phonon-induced backscattering transitions (with $\nu'=-\nu$) become possible because of 
$H'_{\rm e-ph}$ in Eq.~\eqref{newheph}. Fermi's golden rule then yields the transition probabilities
\begin{eqnarray}\label{trans2}
W^\pm_{-\nu,\nu}(p',p) &=& 2\pi v_1^2 Z^2  |2\nu k_F+p-p'| \\ &\times&
\delta \left(-\nu [2k_F+ v_1 (p+p')] \pm \Omega_{p-p'+2\nu k_F} \right).\nonumber
\end{eqnarray}

We now turn to the conductance correction,  $G=G_0+\Delta G(T)$, arising due to phonon-induced backscattering transitions.   We follow Ref.~\cite{Gurevich1995} and consider a TI wire of length $L$ across which 
a small bias voltage $V$ is applied.  The quasi-classical distribution function of fermionic quasiparticles at position $z$ with momentum $\nu k_F + p$ (where $|p|\ll k_F$ and $\nu=\pm$ for right- or left-moving particles) is denoted by $f_\nu(z,p)$. 
For $A=0$, this distribution function reduces to a $z$-independent Fermi-Dirac distribution,
\begin{equation}
\left.    f_\nu(z,p)\right|_{A= 0} = f_\nu^{(0)}(p)\equiv \frac{1}{e^{\nu(v_1p-eV/2)/T}+1}.
\end{equation}
Writing $f_\nu(z,p)=f^{(0)}_\nu(p)+\Delta f_\nu(z,p)$, the Boltzmann equation is given by \cite{Gurevich1995}
\begin{equation}\label{boltzmann}
    \nu v_1  \partial_z \Delta f_\nu = I\left[f^{(0)}\right] + e \partial_z \phi_e \ \partial_p f^{(0)}_\nu,
\end{equation}
where $\phi_e(z)$ is the electrostatic potential along the wire. 
With the shorthand notation $q_\nu=p-p'+2\nu k_F$, the collision integral (omitting the superscripts `$(0)$' in intermediate steps) is given by
\begin{widetext}
\begin{eqnarray}
I\left [f_\nu(p)\right] &=&  - \int \frac{dp'}{2\pi} \Bigl\{ 
W^+_{-\nu,\nu}(p',p)
\left[ f_\nu(p) \left(1-f_{-\nu}(p')\right) (1+N_{q_\nu} )  - f_{-\nu}(p')\left(1-f_\nu(p)\right) N_{q_\nu} \right]  \nonumber\\
& +& W^-_{-\nu,\nu}(p',p)
\left[ f_\nu(p) \left( 1-f_{-\nu}(p')\right) N_{-q_{\nu}}  - f_{-\nu}(p')\left(1-f_\nu(p)\right) (1+N_{-q_\nu})\right] 
\Bigr\},\label{trans4}
\end{eqnarray}
where phonons are distributed according to the Bose-Einstein distribution function,
$N_q = 1/\left(e^{\Omega_q/T}-1\right)$.  
Inserting the transition probabilities \eqref{trans2} into Eq.~\eqref{trans4}, 
we find
\begin{equation}
I\left[f_\nu(p)\right] = -2\sinh \left(\frac{\nu eV}{2T}\right)  f_\nu(p) \int \frac{dp'}{2\pi} 
 f_{-\nu}(p') N_{q_\nu}  \left[ W^+ _{-\nu,\nu}(p',p) e^{\nu v_1 p/T} + 
W^-_{-\nu,\nu}(p',p) e^{-\nu v_1 p'/T} \right] . \label{transf}
\end{equation}
\end{widetext}
By using the identity $W^+_{\nu',\nu}(p',p)=W^-_{\nu,\nu'}(p,p')$, we observe that
$\sum_{\nu=\pm} \int \frac{dp}{2\pi} I\left[f_\nu(p)\right] = 0.$

Solving the Boltzmann equation \eqref{boltzmann} as detailed in Ref.~\cite{Gurevich1995},
the conductance correction then follows as
\begin{equation}\label{corr}
\Delta G = \lim_{V\to 0} \frac{eL}{V} \int \frac{dp}{2\pi} I\left[f^{(0)}_+(p)\right].
\end{equation}
Next we observe that the $\delta$-function in the transition probabilities \eqref{trans2}
enforces the  energy conservation condition $v_1(p'+p)=\pm c_L |2k_F + p-p'|$. Taking into 
account that $c_{\rm L}\ll v_1$ and $|p|,|p'|\ll k_F$, the solution is given by 
 $p' \simeq -p \pm 2k_F c_{\rm L}/v_1$. To lowest order in $V$, Eq.~\eqref{transf} then gives
\begin{eqnarray}\label{transf2}
I\left[f_+(p)\right] &\simeq& - 2k_F v_1 Z^2 N_{2k_F}  \frac{eVe^{T_{\rm BG}/2T}}{T}  \\ \nonumber
&\times&\sum_\pm  f_+(p) \left( 1-f_+(p\mp T_{\rm BG}/v_1) \right) e^{\pm T_{\rm BG}/2T},
\end{eqnarray}
with the Bloch-Gr\"uneisen temperature  $T_{\rm BG}\equiv 2 c_{\rm L} k_F.$
Once $T$ drops below $T_{\rm BG}$, phonon-induced $2k_F$-backscattering becomes suppressed since phonon modes with the required energy of 
order $\Omega_{2k_F}$ are not available anymore.  One then basically has only 
forward scattering processes, where the corresponding transition amplitudes vanish and one therefore expects an 
exponential suppression of the phonon-induced resistivity, see Refs.~\cite{Gurevich1995,Gurevich1995b,Seelig2005}.

Performing the integration in Eq.~\eqref{corr}, the conductance reduction is given by
\begin{equation}
\frac{\Delta G(T) }{e^2/h} = -2 k_F L Z^2 \frac{T_{\rm BG}/(2T)}{\sinh^2\left[T_{\rm BG}/(2T)\right]}.
\end{equation}
As a consequence, the \emph{phonon-induced electrical resistivity} is
\begin{eqnarray}\nonumber
    \rho(T) &=& \frac{h}{e^2} 
    \frac{2 (v_2/v_1)^2}{\pi \rho_M  T_{\rm BG}} \left(\frac{ (1-2\sigma)\alpha\, \delta \Phi_B}{ v_1 R^2 }\right)^2 \times \\ &\times& 
   \frac{T_{\rm BG}/(2T)}{\sinh^2\left[ T_{\rm BG}/(2T)\right]}, \label{resistivity}
\end{eqnarray}
where we have used the definitions of $A$ and $\gamma$ in Eq.~\eqref{Zdef} as well as $k_F\approx\mu/v_1$, see Eq.~\eqref{gammadef}).
At fixed temperature and chemical potential, 
the resistivity thus scales as $\rho\propto (\alpha \delta \Phi_B / R^2)^2$ with the deformation potential coupling $\alpha$, 
the deviation $\delta \Phi_B$ of magnetic flux from the nearest half-integer value, and the nanowire radius $R$. 
In particular the prediction $\rho\propto \delta \Phi_B^2$ may allow for direct tests of our theory using available TI nanowires
 \cite{Ziegler2018,Munning2019}.  
 At low temperatures compared to the Bloch-Gr\"uneisen temperature, Eq.~\eqref{resistivity} implies an exponential suppression of the resistivity, 
$\rho(T\ll T_{\rm BG})\propto (T_{\rm BG}/T) e^{-T_{\rm BG}/T}$, as expected from Refs.~\cite{Gurevich1995,Gurevich1995b,Seelig2005}.
On the other hand, at high temperatures,  the standard linear $T$ dependence, $\rho(T\gg T_{\rm BG})\propto T/T_{\rm BG}$, is recovered.

\section{Conclusions}\label{sec7}

In this work, we have constructed an analytical theory for the coupled electron-phonon system in a topological insulator nanowire pierced by the magnetic flux $\Phi_B$. For half-integer $\Phi_B$, the electronic surface states are represented by topologically protected helical Dirac fermions, where phonons cannot induce $2k_F$ backscattering.  A non-vanishing phonon-induced resistivity emerges only when one has a finite  deviation $\delta \Phi_B$ from half-integer flux values,
where we give detailed predictions for the dependence of the resistivity on temperature and on $\delta \Phi_B$.
 We have also shown that the phase diagram for half-integer flux contains a significant region where superconducting pairing of the surface states is possible.
Future theoretical work could analyze the resistivity for finite $\delta \Phi_B$ taking into account electron-electron interactions ($K\ne 1$), where the approach
of Ref.~\cite{Seelig2005} should allow for progress.  Another interesting avenue for further research concerns the multi-channel generalization of our theory. This case will become important, for instance, at large values of the TI nanowire radius $R$, where our assumption of a single relevant subband breaks down.
 
\begin{acknowledgements}
We acknowledge support by the Deutsche Forschungsgemeinschaft (Bonn) under Grant No.~EG 96/12-1 and under Germany's Excellence Strategy - Cluster of Excellence Matter
and Light for Quantum Computing (ML4Q) EXC 2004/1 - 390534769.
\end{acknowledgements}


\begin{thebibliography}{99}
\bibitem{Fu2007}
L. Fu and C. L. Kane, Phys. Rev. B {\bf 76}, 045302 (2007).

\bibitem{Qi2008}
X. L. Qi, T. L. Hughes, and S. C. Zhang, Phys. Rev. B {\bf 78}, 195424 (2008).

\bibitem{Hasan2010}
M. Z. Hasan and C. L. Kane, Rev. Mod. Phys. {\bf 82}, 3045 (2010).

\bibitem{Liu2010}
C. X. Liu, X. L. Qi, H. J. Zhang, X. Dai, Z. Fang, and S. C. Zhang, Phys. Rev. B {\bf 82}, 045122 (2010).

\bibitem{Qi2011}
X. L. Qi and S. C. Zhang, Rev. Mod. Phys. {\bf 83}, 1057 (2011).

\bibitem{Ando2013}
Y. Ando, J. Phys. Soc. Jpn., {\bf 82}, 102001 (2013).

\bibitem{Zhu2011}
X. Zhu, L. Santos, R. Sankar,  S. Chikara, C. Howard, F. C. Chou,  C. Chamon, and M. El-Batanouny, Phys. Rev. Lett. {\bf 107}, 186102 (2011).

\bibitem{Hatch2011}
R. C. Hatch, M. Bianchi, D. Guan, S. Bao, J. Mi, B. B. Iversen, L. Nilsson, L. Hornekaer, and P. Hofmann, Phys. Rev. B {\bf 83}, 241303(R) (2011).

\bibitem{Zhu2012}
X. Zhu, L. Santos, C. Howard, R. Sankar, F. C. Chou, C. Chamon, and M. El-Batanouny, Phys. Rev. Lett. {\bf 108}, 185501 (2012).

\bibitem{Wang2012}
Y. H. Wang, D. Hsieh, E. J. Sie, H. Steinberg, D. R. Gardner, Y. S. Lee, P. Jarillo-Herrero, and N. Gedik, Phys. Rev. Lett. {\bf 109}, 127401 (2012). 

\bibitem{Kim2012}
D. Kim, Q. Li, P. Syers, N. P. Butch, J. Paglione, S. Das Sarma, and M. S. Fuhrer, Phys. Rev. Lett. {\bf 109}, 166801 (2012).

\bibitem{Pan2012}
Z.-H. Pan, A. V. Fedorov, D. Gardner, Y. S. Lee, S. Chu, and T. Valla, Phys. Rev. Lett. {\bf 108}, 187001 (2012). 

\bibitem{Pan2013}
Z.-H. Pan, E. Vescovo, A. V. Fedorov, G. D. Gu, and T. Valla,
Phys. Rev. B {\bf 88}, 041101(R) (2013). 

\bibitem{Chen2013}
C. Chen, Z. Xie, Y. Feng, H. Yi, A. Liang, S. He, D. Mou, J. He, Y. Peng, X. Liu, Y. Liu, L. Zhao, G. Liu, X. Dong, J. Zhang, L. Yu, X. Wang, Q.
Peng, Z. Wang, S. Zhang, F. Yang, C. Chen, Z. Xu, and X. J. Zhou,  Sci. Rep. {\bf 3}, 2411 (2013). 

\bibitem{Kondo2013}
T. Kondo, Y. Nakashima, Y. Ota, Y. Ishida, W. Malaeb, K. Okazaki, S. Shin, M. Kriener, S. Sasaki, K. Segawa, and Y. Ando, 
Phys. Rev. Lett. {\bf 110}, 217601 (2013). 

\bibitem{Crepaldi2013}
A. Crepaldi, F. Cilento, B. Ressel, C. Cacho, J. C. Johannsen, M. Zacchigna, H. Berger, Ph. Bugnon, C. Grazioli, I. C. E. Turcu, E. Springate, K. Kern, M. Grioni, and F. Parmigiani,
Phys. Rev. B {\bf 88}, 121404(R) (2013).  

\bibitem{Howard2014}
C. Howard and M. El-Batanouny, Phys. Rev. B {\bf 89}, 075425 (2014). 

\bibitem{Costache2014}
 M. V. Costache, I. Neumann, J. F. Sierra, V. Marinova, M. M. Gospodinov, S. Roche, and S. O. Valenzuela, Phys. Rev. Lett. {\bf 112}, 086601 (2014). 

\bibitem{Sobota2014}
J. A. Sobota, S.-L. Yang, D. Leuenberger, A. F. Kemper, J. G. Analytis, I. R. Fisher, P. S. Kirchmann, T. P. Devereaux, and Z.-X. Shen, Phys. Rev. Lett. {\bf 113}, 157401 (2014).

\bibitem{Ando2014} 
Y. Ando, T. Hamasaki, T. Kurokawa, K. Ichiba, F. Yang, M. Novak, S. Sasaki, K. Segawa, Y. Ando, and M. Shiraishi, Nano Lett. {\bf 14}, 6226 (2014).

\bibitem{Glinka2015}
Yu. D. Glinka, S. Babakiray, T. A. Johnson, M. B. Holcomb, and D. Lederman, J. Appl. Phys. {\bf 117}, 165703 (2015).

\bibitem{Tamtogl2017}
A. Tamt\"ogl, P. Kraus, N. Avidor, M. Bremholm, E. M. J. Hedegaard, B. B. Iversen, M. Bianchi, P. Hofmann, J. Ellis, W. Allison, G. Benedek, and W. E. Ernst, Phys. Rev. B {\bf 95}, 195401 (2017). 

\bibitem{Jia2017}
X. Jia, S. Zhang, R. Sankar, F. C. Chou, W. Wang, K. Kempa, E. W. Plummer, J. Zhang, X. Zhu, and J. Guo, Phys. Rev. Lett. {\bf 119}, 136805 (2017).

\bibitem{Wiesner2017} 
M. Wiesner, A. Trzaskowska, B. Mroz, S. Charpentier, S. Wang, Y. Song, F. Lombardi, P. Lucignano, G. Benedek, D. Campi, M. Bernasconi, F. Guinea, and A. Tagliacozzo, Sci. Rep. {\bf 7}, 16449 (2017).  

\bibitem{Huang2008}
B.L. Huang and M. Kaviany, Phys. Rev. B {\bf 77}, 125209 (2008).  

\bibitem{Thalmeier2011}
P. Thalmeier, Phys. Rev. B  {\bf 83}, 125314 (2011).

\bibitem{Giraud2011}
S. Giraud and R. Egger,  Phys. Rev. B {\bf 83}, 245322 (2011).

\bibitem{Giraud2012}
S. Giraud, A. Kundu, and R. Egger, Phys. Rev. B {\bf 85},  035441 (2012).

\bibitem{Budich2012}
J. C. Budich, F. Dolcini, P. Recher, and B. Trauzettel, Phys. Rev. Lett. {\bf 108}, 086602 (2012);
S. Groenendijk, G. Dolcetto, and T. L. Schmidt,
Phys. Rev. B  {\bf 97}, 241406(R) (2018).

\bibitem{Garate2013}
I. Garate, Phys. Rev. Lett. {\bf110}, 046402 (2013).

\bibitem{Zhang2013}
P. Zhang and M. W. Wu, Phys. Rev. B {\bf 87}, 085319 (2013).

\bibitem{Parente2013}
V. Parente, A. Tagliacozzo, F. von Oppen, and F. Guinea, Phys. Rev. B {\bf 88}, 075432 (2013).

\bibitem{Sarma2013}
S. Das Sarma and Q. Li, Phys. Rev. B {\bf 88}, 081404(R) (2013).

\bibitem{Wang2014}
M. Q. Weng and M. W. Wu, Phys. Rev. B {\bf 90}, 125306 (2014).

\bibitem{Heid2017}
R. Heid, I. Yu. Sklyadneva, and E. V. Chulkov, Sci. Rep.  {\bf 7}, 1095 (2017). 

\bibitem{Ran2009}
Y. Zhang, Y. Ran, and A. Vishwanath, Phys. Rev. B {\bf 79}, 245331 (2009).

\bibitem{Ostrovsky2010}
P. M. Ostrovsky, I. V. Gornyi, and A. D. Mirlin, Phys. Rev. Lett. {\bf 105}, 036803 (2010).

\bibitem{Zhang2010}
Y. Zhang and A. Vishwanath, Phys. Rev. Lett. {\bf 105}, 206601 (2010).

\bibitem{Bardarson2010}
J. H. Bardarson, P. W. Brouwer, and J. E. Moore, Phys. Rev. Lett. {\bf 105}, 156803 (2010).

\bibitem{Egger2010} 
R. Egger, A. Zazunov, and A. L. Yeyati, Phys. Rev. Lett. {\bf 105}, 136403 (2010).

\bibitem{Kundu2011}
A. Kundu, A. Zazunov, A. L. Yeyati, T. Martin, and R. Egger, Phys. Rev. B {\bf 83}, 125429 (2011).

\bibitem{Bardarson2013}
J.H. Bardarson and J.E. Moore, Rep. Prog. Phys. {\bf 76}, 056501 (2013).

\bibitem{Landau7}
L. D. Landau and E. M. Lifshitz, \textit{Theory of Elasticity} (Elsevier, 1986).

\bibitem{Svizhenko1998}
A. Svizhenko, A. Balandin, S. Bandyopadhyay, and M. A. Stroscio, Phys. Rev. B {\bf 57}, 4687 (1998).

\bibitem{Dufouleur2013} 
J. Dufouleur, L. Veyrat, A. Teichgr\"aber, S. Neuhaus, C. Nowka, S. Hampel, J. Cayssol, J. Schumann, B. Eichler, O. G. Schmidt, B. B\"uchner, and R. Giraud, Phys. Rev. Lett. {\bf 110}, 186806 (2013).

\bibitem{Hong2014}
S. S. Hong, Y. Zhang, J. J. Cha, X. L. Qi, and Y. Cui, Nano Lett. {\bf 14}, 2815 (2014).

\bibitem{Cho2015}
S. Cho, B. Dellabetta, R. Zhong, J. Schneeloch, T. Liu, G. Gu, M. J. Gilbert, and N. Mason, Nat. Comm. {\bf 6}, 7634 (2015).

\bibitem{Ziegler2018}
J. Ziegler, R. Kozlovsky, C. Gorini, M.-H. Liu, S. Weish\"aupl, H. Maier, R. Fischer, D. A. Kozlov, Z. D. Kvon, N. Mikhailov, S. A. Dvoretsky, K. Richter, and D. Weiss, Phys. Rev. B {\bf 97}, 035157 (2018).

\bibitem{Munning2019}
F. M\"unning, O. Breunig,  H. F. Legg, S. Roitsch, D. Fan, M. R\"o{\ss}ler, A. Rosch, and Y. Ando, arXiv:1910.07863.

\bibitem{Capper2011} 
P. Capper, J. Garland, S. Kasap, and A. Willoughby, {\it Mercury Cadmium Telluride: Growth, Properties and Applications} (Wiley Series in Materials for Electronic and Optoelectronic Applications,
Wiley, 2011).

\bibitem{Jain2013}
A. Jain, S. P. Ong, G.-R. Hautier, W. Chen, W. D. Richards, S. Dacek, S. Cholia, D. Gunter, D. Skinner, G. Ceder, and K. A. Persson, APL Materials {\bf 1}, 011002 (2013).

\bibitem{Gogolinbook}
A. O. Gogolin, A. A. Nersesyan, and A. M. Tsvelik, \textit{Bosonization and Strongly Correlated Systems} (Cambridge University Press, Cambridge UK, 1998).

\bibitem{Pal2012}
H. K. Pal, V. I. Yudson, and D. L. Maslov, Phys. Rev. B {\bf 85}, 085439 (2012).  

\bibitem{Loss1992}
D. Loss and T. Martin, Phys. Rev. B {\bf 50}, 12160 (1994).

\bibitem{DeMartino2003}
A. De Martino and R. Egger, Phys. Rev. B {\bf 67}, 235418 (2003).

\bibitem{Schulz2010}
A. Schulz, A. De Martino, and R. Egger, Phys. Rev. B {\bf 82}, 033407 (2010).

\bibitem{Cook2011}
A. Cook and M. Franz, Phys. Rev. B {\bf 84}, 201105(R) (2011).

\bibitem{Cook2012}
A. M. Cook, M. M. Vazifeh, and M. Franz, Phys. Rev. B {\bf 86}, 155431 (2012).

\bibitem{Fidkowski2011}
L. Fidkowski, R. M. Lutchyn, C. Nayak, and M. P. A. Fisher, Phys. Rev. B {\bf 84}, 195436 (2011).

\bibitem{Manousakis2017}
J. Manousakis, A. Altland, D. Bagrets, R. Egger, and Y. Ando, Phys. Rev. B {\bf 95}, 165424 (2017).

\bibitem{Voit1987}
J. Voit and H. J. Schulz, Phys. Rev. B {\bf 34}, R7429 (1986).

\bibitem{Bockelmann1990}
U. Bockelmann and G. Bastard, Phys. Rev. B {\bf 42}, 8947 (1990).

\bibitem{Shik1993}
A. Y. Shik and L. J. Challis, Phys. Rev. B {\bf 47}, 2082 (1993).

\bibitem{Mickevicius1993}
R. Mickevicius and V. Mitin, Phys. Rev. B {\bf 48}, 17194 (1993).

\bibitem{Gurevich1995}
V. L. Gurevich, V. B. Pevzner, and K. Hess, Phys. Rev. B {\bf 51}, 5219 (1995).

\bibitem{Gurevich1995b}
V. L. Gurevich, V. B. Pevzner, and E. W. Fenton, Phys. Rev. B {\bf 51}, 9465 (1995).

\bibitem{Seelig2005}
G. Seelig, K. A. Matveev, and A. V. Andreev, Phys. Rev. Lett. {\bf 94}, 066802 (2005).

\bibitem{Yurkevich2013}
I. V. Yurkevich, A. Galda, O. M. Yevtushenko, and I. V. Lerner, Phys. Rev. Lett. {\bf 110}, 136405 (2013).

\bibitem{Pochhammer1876}
L. Pochhammer, Journal f\"ur die reine und angewandte Mathematik (Crelle) {\bf 81}, 324 (1876).

\bibitem{Chree1889}
C. Chree, Trans. Camb. Phil. Soc. {\bf 14}, 250 (1889).

\bibitem{Love}
A. E. H. Love, \textit{A Treatise on The Mathematical Theory of Elasticity}
(Dover Publications, New York, 1944).

\bibitem{Graff}
K.F. Graff, \textit{Wave Motion in Elastic Solids} (Dover Publications, New York, 2012).

\bibitem{Jenkins1972}
J.O. Jenkins, J. A. Rayne, and R. W. Ure, Jr., Phys. Rev. B {\bf 5}, 3171 (1972).

\bibitem{DLMF}
F. W. J. Olver, A. B. Olde Daalhuis, D. W. Lozier, B. I. Schneider, R. F. Boisvert, C. W. Clark, B. R. Miller, and B. V. Saunders (eds.), \textit{NIST Digital Library of Mathematical Functions}, available at http://dlmf.nist.gov/, Release 1.0.16 of 2017-09-18. 

\bibitem{Sirenko1997} 
Y. M. Sirenko, K. W. Kim, and M. A. Stroscio, Phys. Rev. B {\bf 56},  15770 (1997).

\bibitem{footnew}
The angular dependence of the superconducting order parameter may include a phase winding factor $\propto e^{in\phi}$ (with some integer $n$) due to the magnetic flux, see
Ref.~\cite{Lutchyn2019a} for a related case.

\bibitem{Lutchyn2019a}
R. M. Lutchyn, G. W. Winkler, B. van Heck, T. Karzig, K. Flensberg, L. I. Glazman, and C. Nayak, arXiv:1809.05512.

\bibitem{Kurilo1997}
I. V. Kurilo, V. P. Alekhin, I. O. Rudyi, S. I. Bulychev, and L. I. Osypyshin, phys. stat. sol. (a), {\bf 163}, 4758 (1997).



\end{thebibliography}
\end{document}